\documentclass[letter]{article}

\usepackage[english]{babel}
\usepackage[utf8x]{inputenc}
\usepackage[T1]{fontenc}
\usepackage{setspace}

\usepackage[sort&compress,square,numbers]{natbib}

\usepackage[a4paper,top=3cm,bottom=2cm,left=3cm,right=3cm,marginparwidth=1.75cm]{geometry}

\usepackage{amsmath}
\usepackage{amsfonts}
\usepackage{amssymb}
\usepackage{amsthm}
\usepackage{graphicx}
\usepackage{authblk}
\usepackage[colorinlistoftodos]{todonotes}
\usepackage[colorlinks=true, allcolors=blue]{hyperref}

\newcommand{\bsym}{\boldsymbol}

\newtheorem{defn}{Definition}

\title{Evaluating the Success of a Data Analysis}
\author[1]{Stephanie C. Hicks\footnote{Corresponding author email: \href{shicks19@jhu.edu}{shicks19@jhu.edu}}}
\author[1]{Roger D. Peng}
\affil[1]{Department of Biostatistics, Johns Hopkins Bloomberg School of Public Health}

\begin{document}
\maketitle

\doublespacing

\begin{abstract}

A fundamental problem in the practice and teaching of data science is how to evaluate the quality of a given data analysis, which is different than the evaluation of the science or question underlying the data analysis. Previously, we defined a set of principles for describing data analyses that can be used to create a data analysis and to characterize the variation between data analyses. Here, we introduce a metric of quality evaluation that we call the \textit{success} of a data analysis, which is different than other potential metrics such as completeness, validity, or honesty. We define a successful data analysis as the matching of principles between the analyst and the audience on which the analysis is developed. In this paper, we propose a statistical model and general framework for evaluating the success of a data analysis. We argue that this framework can be used as a guide for practicing data scientists and students in data science courses for how to build a successful data analysis. 

\end{abstract}t


\noindent \textbf{Running title}: Evaluating the success of a data analysis

\noindent \textbf{Author Contributions}: SCH and RDP equally conceptualized, wrote and approved the manuscript. 

\noindent \textbf{Disclosures}: The authors do not have any disclosures. 

\noindent \textbf{Acknowledgements}: The authors do not have any funding to acknowledge.

\clearpage

\section{Introduction}

Within the practice and teaching of data science \cite{Cleveland2001, nolanlang2010, asaundergrad2014, baumer2015, pwc-datascience, hardin2015, hortonhardin2015, Donoho2017, kaplan2018, hicks2018}, a data scientist builds a data analysis  \cite{tuke:1962, tukeywilk1996, box1976, wild1994, chatfield1995, wildpfannkuch1999, cook2007} to extract knowledge and insights from examining data \citep{grol:wick:2014}. However, there is surprisingly little discussion on how to evaluate the quality of a given data analysis, which is different than the evaluation of the science or question underlying the data analysis. Three possible reasons for this include (1) there is an insufficient vocabulary to describe how to characterize the variation between data analyses, (2) there is a lack of definitive and precise performance metrics to evaluate the quality of the analyses, and (3) there is lack of specificity by whom the data analysis is being evaluated. This leaves the educator or the practicing data scientist to focus the discussion of data analysis quality assessment on specific methods, technologies or programming languages used in a data analysis, with the vague hope that such discussion will lead to success. 

Much previous work dedicated to studying data analysis has focused primarily on the notion of ``statistical thinking'', or developing an understanding of the mental processes that occur within the analyst while doing data analysis~\citep{wildpfannkuch1999,grol:wick:2014,wats:call:2003, hortonhardin2015}. Such an approach is beneficial in that by understanding how data analyses are conceived we can design teaching strategies that are purpose-built to emphasize certain processes. An alternate approach is to characterize the data analytic process based on its observed outputs---the data analysis---and provide principled feedback on why it might have failed or how it could be more successful. However, the literature provides little insight into how we might execute this approach, largely because there is no rigorous description of a ``successful'' data analysis.

The current situation leads us to re-think the purpose of a data analysis and the \textit{audience} that it serves. While the audience could be one individual, or a group of individuals, each individual audience member plays a critical role in evaluating the quality of a given data analysis. Each audience member evaluates the quality with her or his own preconceived notions, characteristics, and biases towards valuing what makes a good or bad analysis~\citep{wildpfannkuch1999}. Therefore, to be able to define precise performance metrics to evaluate the quality of a data analysis, we first need to formally specify (i) who is the audience and (ii) what characteristics do they value, or not value, in a given data analysis. With this information in hand, a data analyst could then hypothetically choose to adjust or tailor a given data analysis to the characteristics that the audience members value, leading to potentially a more \textit{successful} data analysis, compared to one that did not take into account the audience and the characteristics in a data analysis that the audience members value~\citep{peng:2018,peng:2018a}. 

In contrast, there are other potential metrics of quality evaluation one could consider, such as whether or not an analysis is valid or complete, or even evaluating the strength and quality of evidence in a given data analysis for the particular hypothesis of interest \citep{hickspeng2019-elementsandprinciples}. While all of these quality evaluations of data analyses are important, in this paper, we are focused on the question of how to evaluate the \textit{success} of a data analysis, which will depend on formally specifying who is the audience and the characteristics in a data analysis that the audience members value. 

To tackle this question, we start by leveraging a set of principles of data analyses that we previously introduced that can be used to create a data analysis and to characterize the variation between data analyses. These principles of data analysis are ``prioritized qualities or characteristics that are relevant to the analysis, as a whole or individual components, and that can be objectively observed or measured'' \citep{hickspeng2019-elementsandprinciples}. For a given data analysis, the inclusion or exclusion of certain principles does not convey a judgment or assessment with respect to the overall quality of the data analysis. However, a data analyst can assign weights to these principles to increase or decrease the presence of these objective characteristics in a given data analysis, which can be also highly influenced by outside constraints or resources, such as time or budget. In this way, different weighting of the principles by the analyst can lead to different data analyses, all addressing the same primary question underlying the data analysis~\citep{Silberzahn2018}. 

Next, we use this set of principles for data analysis to propose a framework for evaluating the quality of a data analysis that relies critically on the audience for which the analysis is developed. In particular, as every data analysis has an audience that views the analysis with her or his own preconceived notions, characteristics, and biases, we consider the weights of the principles by both the analyst and the audience members, who may have a different perspective of how these various principles should be weighted for a given data analysis. For example, one audience (Audience A) may value one set of principles while another audience (Audience B) may value a different set of principles. Neither set of principles weighted by the analyst or either set of audiences, is correct or incorrect. However, we previously hypothesized that the success of a data analysis may depend on how well-matched the analyst's weightings are to the audience's weightings for a given analysis~\citep{hickspeng2019-elementsandprinciples}. In this way, educators can use this idea in the classroom to teach students how to build more successful analyses that take into account who is the audience and what principles of data analysis that they value. In addition, managers of data analysts in industry can use this idea to frame the discussion of how to to build more successful data analyses for their clients, customers, or executives. 

In this paper, we make these ideas more concrete and introduce a metric of quality evaluation that we call the \textit{success} of a data analysis. We define a successful data analysis as the matching of weighted principles between the analyst and the audience on which the analysis is developed. In the following sections, we mathematically formalize those ideas by proposing a statistical model and general framework for evaluating the success of a data analysis (Section \ref{sec:framework}). Then, we discuss the implications of this framework (Section \ref{sec:implications}) and argue how this framework can be used as a guide for practicing data scientists and students in data science courses for how to build a successful data analysis.

\section{Statistical Framework for Evaluating Data Analyses}
\label{sec:framework}

As described above and our in previous work, we consider data analyses to be constructed in a manner guided by a set of $K$ principles \citep{hickspeng2019-elementsandprinciples} or objective characteristics about the data analysis. Specifically, we defined the principles of data analysis as \textit{data-matching}, \textit{exhaustive}, \textit{skeptical}, \textit{second-order}, \textit{transparent}, and \textit{reproducible}. In this paper, we assume that for each principle, the data analyst assigns a positive integer score whose interpretation corresponds to how much weight that individual gives to that principle. We consider smaller values to be interpreted as a ``lower weight'' assigned to a given principle and larger values interpreted as a ``higher weight''. In some circumstances, it may make sense to think of the weight as the number of units of a particular resource, such as time or budget, that is devoted to a given principle.

For a given data analysis, an analyst will assign a weight $W^{(k)}$ to principle $k$. For example, if we assume principle $k$ is reproducibility, the analyst might assign a weight $W^{(k)} = 100$ to a data analysis because the analyst believes reproducibility is very important for that analysis. For a different analysis, where the reproducibility of the results is perhaps not so critical, the analyst may assign a weight $W^{(k)} = 10$ for this specific principle. Given a set of $K$ principles, an analyst assigns a set of weights $\left\{W^{(1)},\dots,W^{(K)}\right\}\in\mathbb{Z}_+^K$ to guide the development of this data analysis. The sets of weights assigned to each principle may differ from analysis to analysis.

Data analyses are built to be viewed by an audience, which can be an individual person or a group of people and can include the data analyst them self. For now, we will consider the audience to be an individual person, other than the data analyst, and consider the case when an audience is more than one individual in Section \ref{sec:group-audiences}. As such, the audience has their own weights for each principle governing a data analysis, which reflects how they balance the importance of various properties of a data analysis. For a given data analysis, the audience weights will be denoted by the set $\left\{A^{(1)},\dots,A^{(K)}\right\}\in\mathbb{Z}_+^K$. These values are assigned before seeing the full data analysis, but may be based on partial information available about the analysis or analyst beforehand.

\subsection{Fixed Variation in Weightings}

We allow for the possibility that there will be variation in the weightings of the principles from analysis to analysis, for both analyst and audience. Some of that variation can be characterized as fixed, while other variation may be best considered as random. From the analyst's perspective, some of the determinants of how a given principle may be weighted are:
\begin{enumerate}
    \item \textit{Analysis-specific Resources}. Considerations about computing resources, time, budget, personnel, and other such resources and analysis characteristics can often require that an analyst place more or less weight on certain principles for analysis. For example, analyses that must be conducted in a short amount of time may be limited in their ability to explore multiple competing hypotheses and exhibit low skepticism.
    \item \textit{Question Significance and Problem Characteristics}. The significance of the question being addressed with the data may play a role in determining principle weightings. Questions of high significance, for example, may require a high degree of transparency or reproducibility. Questions of lower significance may be done in a ``quick-and-dirty'' fashion; should the question's significance change in the future the analysis may need to be re-done with a different set of principle weightings.
    \item \textit{Field-specific Conventions}. Analysts are often members of a field from which they may have received their training (e.g. statistics, economics, computer science, bioinformatics). Each field develops conventions regarding how analyses in their field should be conducted and we characterize this using a field-specific mean value for a given principle. Tukey~\cite{tuke:1962} emphasized that in data analysis, there is a heavy emphasis on ``judgment'', one particular form of which is based upon the experience of members of a given field.
    \item \textit{Analytic Product}. Depending on the analytic product that will ultimately be presented to the audience (e.g. PDF document, web-based dashboard, executable R Markdown document), the analyst may determine that certain principles should receive more or less weight. 
\end{enumerate}

Similarly, the audience for whom the analysis is being developed will determine their principle weightings based on a variety factors, including their \textit{perception} of resources available to the analyst, their judgment of the significance of the question, their own field-specific conventions (assuming the audience and the analyst are not members of the same field), and their perception of what the analytic product should contain.

\subsection{Random Variation in Weightings}

The above-enumerated list describes some of the fixed factors that may drive variation in how various data analytic principles are weighted. However, there may be variation that is more random in nature. In particular, we consider the randomness as arising from sampling from a population of analysts or potential audience members. Different analysts, presented with the exact same question and data, will likely weight principles differently and hence produce different analyses based on their own personal characteristics. Similarly, different audience members, seeing the same analytic product, will weight principles differently and evaluate the success of the analysis differently.

We consider each analyst and each audience member to be a member of a \textit{field} or profession. Let $f_i\in\{1,\dots,F\}$ be the index into a set of $F$ fields or professions for analyst $i$. One source of random variation that we highlight here is what we call an individual's \textit{field-specific deviation} for principle~$k$. An analyst who belongs to field $f_i$ will be trained in the conventions of that field, which places a field-specific mean value $\lambda^{(k)}_{f_i}$ for a given principle~$k$. An individual analyst~$i$ will deviate from their field-specific mean by an amount $\delta_i^{(k)}$ which we think of as being randomly distributed with mean $0$ and finite variance. Therefore, the field-specific principle contribution for analyst~$i$ is $\lambda^{(k)}_{f_i} + \delta^{(k)}_{i}$ for principle $k$ in any given data analysis. Similarly, audience member $j$ who belongs to field $f_j$ will have a field-specific principle contribution of $\lambda^{(k)}_{f_j} + \eta^{(k)}_{j}$, where $\eta^{(k)}_{j}$ is randomly distributed with mean $0$ and finite variance.

\subsection{Model for Principle Weights}

Throughout text, we consider just one data analysis $a$ at a time, but we do not include the notation for the $a^{th}$ data analysis to keep the notation minimal. Now, for a given analysis and analyst $i$, the weight assigned to a specific principle $k$ is $W_{i}^{(k)}$ and $N_i = \sum_{k=1}^K W_{i}^{(k)}$ is the total weight assigned to the analysis by analyst $i$. Given the total weight $N_{i}$, we model the individual principle-specific weights $W_{i}^{(k)}$ with the multinomial distribution,
\begin{equation}
\mathbf{W}_i 
=
\left(W_{i}^{(1)},\dots,W_{i}^{(K)}\right) 
\sim
\text{Multinomial}\left(N_{i}; \pi^{(1)}_{i},\dots,\pi^{(K)}_{i}\right).
\label{eq:analyst}
\end{equation}
The parameters $\pi_{i}^{(k)}$ from the multinomial distribution can be thought of as the probability of analyst $i$ assigning weight to a specific principle $k$ where the probabilities must sum to 1 across the $K$ principles, i.e. $\sum_{k=1}^K \pi_{i}^{(k)} = 1$, reflecting the reality that all analysts must decide how to allocate their priorities towards each principle when building a data analysis. For a given principle $k$, we can derive the marginal distribution from the multinomial and have
$$
W_{i}^{(k)}
\sim
\text{Binomial}(N_{i}; \pi_{i}^{(k)}).
$$
We can then model the $\pi_{i}^{(k)}$s as
\begin{equation}
\psi_{i}^{(k)}
=
\log\left(\frac{\pi_{i}^{(k)}}{1-\pi_{i}^{(k)}}\right)
=
\lambda_{f_i}^{(k)} + \delta_{i}^{(k)} + \mathbf{x}_i^\prime\bsym{\beta}^{(k)}_i,
\end{equation}
where $\lambda_{f_i}^{(k)}$ is the field-specific mean for principle $k$ and analyst $i$ in the field $f_i$, $\delta_{i}^{(k)}$ is analyst $i$'s deviation from the field-specific mean for principle $k$, $\mathbf{x}_i$ is a vector of analysis-specific resources and characteristics for the analysis (i.e. time, budget, personnel, significance), and $\bsym{\beta}^{(k)}_i$ is a vector of coefficients that indicate how each resource is related to the up-weighting or down-weighting of the $k^{th}$ principle for this analysis. We consider the analyst deviation $\delta_{i}^{(k)}$ to be randomly distributed across the set of potential analysts with mean $0$ and finite variance.

Analogous to the analyst's weights, the weight given to principle $k$ by audience member $j$ (who is a member of field $f_j$) can be written as $A_{j}^{(k)}$ with $N_{j}=\sum_{k=1}^K A^{(k)}_{j}$ being the total weight given to the analysis. We similarly model the vector $\mathbf{A}_j = 
\left(A_{j}^{(1)},\dots,A_{j}^{(K)}\right)$ as multinomial with total $N_{j}$ and proportions $\omega_{j}^{(1)},\dots,\omega_{j}^{(K)}$.
We then similarly model the proportions $\omega_{j}^{(k)}$ as
\begin{equation}
\alpha^{(k)}_{j}
=
\log\left(\frac{\omega^{(k)}_{j}}{1-\omega^{(k)}_{j}}\right)
=
\lambda^{(k)}_{f_j} + \eta^{(k)}_{j} + \mathbf{z}_j^\prime\bsym{\gamma}^{(k)}_j
\end{equation}
where $\mathbf{z}_j$ is the audience's perception of resources available and question significance, $\lambda^{(k)}_{f_j}$ and $\eta_{j}^{(k)}$ are the field-specific mean and individual-level deviation for the $j^{th}$ audience member, respectively, and $\bsym{\gamma}_j^{(k)}$ is the audience member's sense of the relationship between a given resource and the weight that should be given to the principle. Note that we consider $\eta_{j}^{(k)}$ to be independent of $\delta_{i}^{(k)}$ in the analyst's weight model.

With the analyst weightings in Equation~(\ref{eq:analyst}) and the audience weightings, we can then write the principle-specific weight difference for a given data analysis as
\begin{eqnarray}
D_{ij}^{(k)}
& = &
\psi^{(k)}_{i} - \alpha^{(k)}_{j}\nonumber \\
& = &
\left(\lambda_{f_i}^{(k)} - \lambda_{f_j}^{(k)}\right) + \left(\delta_{i}^{(k)}-\eta_{j}^{(k)}\right)
+ \left(\mathbf{x}_i^\prime\bsym{\beta}^{(k)}_i - \mathbf{z}_j^\prime\bsym{\gamma}^{(k)}_j\right)
\label{eq:distance}
\end{eqnarray}
The overall analyst-audience distance for a given data analysis is then characterized by the collection of distances for the set of $K$ principles $\mathbf{D}_{ij}=\left(D_{ij}^{(1)},\dots,D_{ij}^{(K)}\right)$. 

In the next section, we will introduce three ways that a given data analysis can be defined as successful. 

\subsection{Defining a Successful Data Analysis}

In this section, we propose three ways to achieve a successful data analysis pairwise between the analyst $i$ and audience member $j$: \textit{Strong Pairwise Success} (Definition 1), \textit{Weak Pairwise Success} (Definition 2), \textit{Potential Pairwise Success} (Definition 3). 

\begin{defn}[Strong Pairwise Success]
A data analysis is strongly successful for the pairing of analyst $i$ with audience member $j$ if
$$
\left\|\mathbf{D}_{ij}\right\|_\infty 
= 
\max_{k=1,\dots,K} \left|D_{ij}^{(k)}\right| < \varepsilon.
$$
\end{defn}
\noindent
for some small $\varepsilon$. Because of the randomness in $\delta_{i}^{(k)}$ and $\eta_{j}^{(k)}$, the $D_{ij}^{(k)}$ values can never be equal to zero. However, the definition of strong pairwise success requires that the differences are never too large for any given principle.

We can propose a weaker form of analysis success that allows for some differences in how the principles are weighted, but places a limit on the total variation of those differences.
\begin{defn}[Weak Pairwise Success]
A data analysis is weakly successful for the pairing of analyst $i$ with audience member $j$ if for some $p\geq 1$
\begin{equation}
\left\|\mathbf{D}_{ij}\right\|_p
=
\left(\frac{1}{K}\sum_{k=1}^K \left|D_{ij}^{(k)}\right|^p\right)^{1/p} < \varepsilon.
\label{eq:lpnorm}
\end{equation}
\end{defn}
\noindent
With this definition, the analyst and audience may differ slightly with respect to how each principle is weighted, but the overall differences between analyst and audience must be small. The choice of $p$ here (and hence, the norm) will have an impact on how much deviation is allowed between analyst and audience and how much any single principle may differ. For now, we do not comment on which norm is most appropriate or useful, but only note that different circumstances may require the use of different norms.

From our definition of strong pairwise success of a data analysis, we can see how success may be achieved or, in some circumstances, may never be achieved. In particular, if we consider $\delta_{i}^{(k)}$ and $\eta_{j}^{(k)}$ to be random (with mean $0$ and finite variance) and independent, then the principle-specific weight difference has expectation
\begin{equation}
\mathbb{E}\left[D_{ij}^{(k)}\right]
=
(\lambda_{f_i}^{(k)} - \lambda_{f_j}^{(k)}) + (\mathbf{x}_i^\prime\bsym{\beta}_i^{(k)} - \mathbf{z}_j^\prime\bsym{\gamma}_j^{(k)}),
\end{equation}
which in general will be different from $0$. 

A separate measure of success can be defined in situations where the analyst $i$ may only have general information about the audience member $j$, but may not know specifically who the audience will be. In such cases, the analyst may have information about the population parameters of the audience and so may wish to measure success based on the mean values for the population. We look at the difference in expected values for the weightings for all $K$ principles and denote this the \textit{potential} pairwise success of an analysis, because we have not yet observed the audience's principle weighting.

\begin{defn}[Potential Pairwise Success]
A data analysis is potentially successful for the pairing of analyst $i$ with audience member $j$ if
$$
\mathbb{E}\left[\mathbf{D}_{ij}\right] = \mathbf{0}.
$$
\end{defn}
\noindent
A key distinction between \textit{strong} (or \textit{weak}) pairwise success and \textit{potential} pairwise success is that the former can only be evaluated when analyst and audience meet and a data analysis is presented. Potential pairwise success can be evaluated before an analyst presents the analysis to the audience. As such, the potential pairwise success metric could serve as a target for optimization by the analyst and we discuss this briefly in the Discussion below.

\subsection{Group Audiences}
\label{sec:group-audiences}
Up until this point we have assumed the audience consisted of a single member indexed by $j$. However, it is common that a data analysis will be reviewed by or presented to a group of audience members. If there are $J$ members of the audience, then we can extend Equation~(\ref{eq:distance}) to be as follows.
\begin{eqnarray}
D_{i\cdot}^{(k)}
& = &
\frac{1}{J}\sum_{j=1}^J D_{ij}^{(k)}\nonumber \\
& = &
\psi^{(k)}_{i} - \frac{1}{J}\sum_{j=1}^J\alpha^{(k)}_{j}\nonumber \\
& = &
\left(\lambda_{f_i}^{(k)} - \frac{1}{J}\sum_j \lambda_{f_j}^{(k)}\right) + \left(\delta_{i}^{(k)}-\frac{1}{J}\sum_j\eta_{j}^{(k)}\right)
+ \left(\mathbf{x}_i^\prime\bsym{\beta}^{(k)}_i - \frac{1}{J}\sum_j\mathbf{z}_j^\prime\bsym{\gamma}^{(k)}_j\right).
\label{eq:distance-group}
\end{eqnarray}
In this formulation, $D_{i\cdot}^{(k)}$ is small if principle $k$ is weighted by the analyst in a manner that is equal to the mean of the members of the audience. With this extension of the principle-specific weight difference to group audiences, we can modify our definition of pairwise potential success to be

\begin{defn}[Potential Group Success]\label{defn:groupsuccess}
A data analysis is potentially successful for analyst $i$ presenting to a group consisting of members $j=1,\dots,J$ if for the vector $\mathbf{D}_{i\cdot} = \left(D_{i\cdot}^{(1)},\dots,D_{i\cdot}^{(K)}\right)$, we have
$$
\mathbb{E}\left[\mathbf{D}_{i\cdot}\right] = \mathbf{0}.
$$
\end{defn}
\noindent Analogous definitions for strong group success and weak group success could be constructed, but we omit them here. We believe the definition of potential group success is the most relevant to data analysts who will be presenting their work to multiple people and may need to consider the heterogeneity of the audience to which they will be presenting.

\section{Implications}
\label{sec:implications}

The definitions of pairwise success and potential pairwise success presented in Section~\ref{sec:framework} lead to several implications about how data analyses may or may not succeed and what could potentially be done to improve the success of any given analysis. We discuss some of these implications in this section. First, it follows from Equation~(\ref{eq:distance}) that one way in which $\mathbb{E}\left[D_{ij}^{(k)}\right]$ could be made to be smaller would be to have the analyst and audience member be from the same field. If analyst $i$ and audience member $j$ have $f_i = f_j$, then we have $\lambda^{(k)}_{f_i} - \lambda^{(k)}_{f_j} = 0$. 

The interpretation of this is that members of the same field share similar conventions with respect to a given principle. For example, if ``computational reproducibility'' is the $k^{th}$ principle, then members of the field of computational biology (for example), which generally places a high weight on computational reproducibility, would on average place a high weight on that principle. We might then expect data analyses in this field to generally demonstrate a high weight on reproducibility, with perhaps code and data routinely made available. As a result, we would expect a higher potential for success (i.e. smaller $\mathbb{E}\left[D_{ij}^{(k)}\right]$) if analyst $i$ and audience member $j$ are both in the field of computational biology.

The random variation in $D_{ij}^{(k)}$ ensures that $D_{ij}^{(k)}$ can never be equal to $0$. In defining strongly successful and weakly successful analyses, we allow for some differences between analyst and audiences. The magnitude of allowable differences in principle weightings, $\varepsilon$, is likely to be analysis-specific and will depend in part on the context and circumstances surrounding the analysis. For a quick, ``work-in-progress'' type of analysis, the audience may allow for larger deviations, with the presumption that the final version will have the appropriate principle weighting. More ``final'' analyses, such as a published paper, may require a stricter adherence to the audience's principle weightings in order to declare success.

In addition, the resources available to the analyst and the specific characteristics of the problem being addressed may lead an analyst to re-prioritize the weights assigned to different principles, leading to a deviation from what they might typically assign based solely on field conventions and personal preference. For analyst $i$, these resources and problem characteristics are denoted by $\mathbf{x}_i$ and the manner in which an analyst re-prioritizes principle $k$ in response to changes in resources or problem characteristics is encoded in the vector $\bsym{\beta}_i^{(k)}$. 

For example if principle $k$ is computational reproducibility and $x_i$ is the time available to the analyst $i$ to do the analysis, then $\beta_i^{(k)}=0$ would imply that analyst places the same amount of weight on this principle no matter how much time is available. However, if principle $k$ is ``exhaustiveness'' and $x_i$ is time available, then $\beta_i^{(k)} > 0$ would imply that the more time that is available for an analysis, the more the analyst prioritizes exhaustiveness relative to the other principles (and similarly, less time available would lead to less weight on exhaustiveness). Furthermore, in our formulation, $\exp\left(\beta_i^{(k)}\right)$ can be interpreted as how many more times principle $k$ is weighted versus all of the other principles under consideration.

The audience's perception of the resources available for conducting the analysis and the problem-specific characteristics of the analysis is encoded in $\mathbf{z}_j$ and can play a role in how different principles are weighted via $\bsym{\gamma}_j^{(k)}$. If $\mathbf{x}_i=\mathbf{z}_j$ then that implies the audience's perception of the resources and characteristics is equal to that of the analyst. If $\bsym{\beta}_i^{(k)}=\bsym{\gamma}_j^{(k)}$ then the audience and analyst have the same understanding of how resources and problem-specific characteristics should affect principle weightings (if at all). 

Ultimately, even if $\mathbb{E}\left[D_{ij}^{(k)}\right]=0$, we can still observe a mismatch between analyst and audience based on individual-level random variation. Each analyst and audience member will randomly deviate from their field-specific mean and the variance of those deviations will play a role in the likelihood of success for a given analysis. If analyst $i$ presents an analysis to audience member $j$ of a field that exhibits wide variation in how they weight a given principle, then the probability of a mismatch is large, even if the analyst's and audience's fields have similar mean values on that principle.

Our definition of potential group success in Definition~\ref{defn:groupsuccess} suggests that an analysts are successful when their individual principle weightings match those of the audience's mean values. If the audience members are are all members of the same field, then we have $\lambda_{f_1}^{(k)}=\cdots=\lambda_{f_J}^{(k)}$, which greatly simplifies any analytic presentation. However, if the audience members are all of different fields, then it may be more difficult to identify the mean of the audience members' $\lambda_{f_j}^{(k)}$ values. Interestingly, as the audience size grows, we should have that $\frac{1}{J}\sum_j \eta_j^{(k)}\rightarrow 0$, regardless of the individual audience members' respective fields, because we assume that an individual audience member's field-specific deviation is random with mean $0$. This suggests that for audiences beyond a certain size, that component of $D_{i\cdot}^{(k)}$ will always be near $0$. This is intuitive as when presenting to a very large audience, it is generally challenging for the analyst to consider the individual needs of every audience member.

\section{Discussion}

In proposing how to define the \textit{success} of a data analysis, our goal is to provide a guide for practicing data scientists and students in data science courses for how to build a successful data analysis. We aim for this to serve as one of many possible performance metrics in the evaluation of data analyses as part of our larger goal for developing a \textit{theory} of data analysis and data science. Other performance metrics could include the \textit{validity} of a data analysis, the evaluation of the \textit{strength of evidence} in the science or question underlying the data analysis, or the \textit{honesty} or \textit{intention} of the data analyst when building a data analysis. 

We have presented the idea of data analysis success in a manner that places the analyst and the audience as essentially passive actors embedded in a larger framework. However, the reality is that the analyst will typically have far more agency in determining the success of an analysis. Furthermore, when the audience is small, there will likely be some communication between analyst and audience, either before the analysis is conducted or while it is ongoing, to better lay the groundwork for success. Such communication could be used to broker agreement on which principles guide the analysis and how each principle is weighted. We have presented the notion of a data analysis as a snapshot of what is more likely a complex dynamical activity, with constant feedback and adjustments taking place over time. 

One important feature that we have not discussed is how can a data analyst $i$ adjust her or his weights (before presenting the analysis) for each principle $k$ based on how they \textit{perceive} the audience would weight each principle. Another way of stating this is that as the analyst $i$ might have partial information about the audience member $j$, such as their background, and other contextual information, the analyst $i$ may choose to adjust their principle weightings based on what they perceive the audience preferences to be. One way to obtain this information would be to directly ask the audience member $j$ about their audience principle-specific weights $A_j^{(k)}$ for a given data analysis. In this way, one could imagine an \textit{audience correction factor} could be added to Equation~(\ref{eq:distance}) to allow for possibility that an analyst might attempt to adjust their analyst principle-specific weighting $W_i^{(k)}$ based on what they expect the audience to prefer. In addition, this correction factor could be a function of the degree by which the analyst attempts to correct for the audience's weighting preferences. In some cases, the analyst will make a strong attempt to adjust the analysis to the audience's preferences and in other cases, the analyst will make minor adjustments, if at all. Developing and characterizing strategies for analysts to actively improve the chances of success is an important area of future work.

Our definition of success in data analysis depends solely on the participants---the analyst and the audience---and the outputs of the data analysis. In theory, one could calculate the pairwise success of an analysis with just those elements. Critically, we do not consider events or information that occur outside the analysis or perhaps in the future. For example, an analysis may make certain conclusions based on the evidence available in the data that are later invalidated by more in-depth analysis (perhaps with better data). We do not therefore conclude that the original analysis was by definition a failure. At any given moment, an analysis can only draw on the data and evidence that are available. It therefore seems inappropriate to judge the success of a data analysis based on information that were not accessible at the time.

Our approach to defining success in data analysis shares many elements with the field of design thinking in its approach to building a solution matched to a specific audience~\citep{cros:2011,peng:park:2018}. In some ways, one could think of a data analysis as a kind of ``product'', in the sense that it is not a naturally occurring object in nature. As such, someone---the analyst---must design the analysis in a manner that makes it useful to the audience. The success of the analysis will depend in part on considering the audience, much like the success of any designed product.

Finally, a benefit of defining success in data analysis is that we can now clearly recognize failure. Learning from failed data analyses is an important aspect of the training of any data analyst and the first step in that process is knowing when failure has occurred. Dialog between audience and analyst about why an analysis has failed can improve the quality of future analyses, as well as improve the quality of the relationship between analyst and audience. Critical to such ``post-mortem'' discussions is that it be conducted in a blameless manner~\citep{park:2017} so that analyst and audience can quickly come to a resolution over how problems should be fixed.

\section{Summary}

The practice and teaching of data science requires the careful evaluation of data analyses. In this paper, we introduce a precise metric of quality evaluation to assess the success of a data analysis. This metric depends on input from both the data analyst and the audience evaluating the data analysis. The benefits of this general framework for evaluating the success of a data analysis include providing a guide to practicing data scientists and students in data science courses on how to build a successful data analysis. 


\clearpage 
\bibliographystyle{unsrtnat}
\bibliography{tods-success.bib}

\end{document}